\documentclass[letter,narrowdisplay]{elsart3p}
\usepackage{graphicx}

\newcommand{\beq}{\begin{equation}}
\newcommand{\eeq}{\end{equation}}
\newcommand{\bea}{\begin{eqnarray}}
\newcommand{\eea}{\end{eqnarray}}
\DeclareMathAlphabet{\mathbbm}{U}{bbm}{m}{n}
\SetMathAlphabet\mathbbm{bold}{U}{bbm}{bx}{n} 

\journal{Physics Letters B}

\begin{document}
\begin{frontmatter}
\begin{flushright}
JLAB-THY-07-640
\end{flushright}
\title{Baryon Regge Trajectories in the Light of the $1/N_c$ Expansion}
\vspace{0.4cm}

\author[hampton,jlab]{J. L. Goity} \ead{goity@jlab.org}$\negthinspace\negthinspace$, \author[jlab,liege]{N.  Matagne} \ead{nmatagne@ulg.ac.be}
\vspace{0.4cm}
\address[hampton]{Department of Physics, Hampton University, Hampton, VA 23668, USA}\address[jlab]{Thomas Jefferson National Accelerator Facility, Newport News, VA 23606, USA}

\address[liege]{University of Li\`ege, Institute of Physics B5, Sart Tilman, B-4000 Li\`ege 1, Belgium}
\date{\today}

\begin{abstract}
We analyze Regge trajectories  in terms of  the $1/N_c$ expansion of QCD. Neglecting spin-orbit contributions to the large $N_c$ baryon mass operator, we consider the evolution of the  spin-flavor singlet component of the masses with respect  to the angular momentum. We find two distinct and remarkably linear Regge trajectories for  symmetric and  for  mixed symmetric spin-flavor multiplets. \\
\end{abstract}

\end{frontmatter}
\maketitle

\section{Introduction}

The ordering of hadronic states on approximately linear Regge trajectories in the Chew-Frautschi plot is one of  the most remarkable features  of the QCD spectrum. It manifests the underlying non-perturbative QCD dynamics,
which at long distances becomes dominated by  the string-like  behavior that leads to confinement.
  In fact this picture has been the motivation for the development of string/flux tube models of hadrons \cite{carlson83}, which contemporarily are described as effective theories  in the so called AdS/QCD  framework \cite{ADSQCD}. The latter is valid in the large $N_c$ limit, $N_c$ being the number of colors, and has been applied  almost exclusively to  mesons, while   extensions   to baryons are being  explored
  \cite{BrodskyTeramond, forkel}. Furthemore, it has been shown recently that flux tube model and large $N_c$ mass formulas are compatible \cite{semay}. Regge trajectories have also been recently considered in the context of the quark-diquark picture of baryons \cite{Wilczek}.
  
In this work we will analyze  the baryon Regge trajectories in the light of the $1/N_c$ expansion, which is in principle an approach consistent with QCD.
The $1/N_c$ expansion for baryons is based on the emergent  $SU(6)$ spin-flavor symmetry (for three light flavors)  in the large   $N_c$ limit  \cite{Gervais:1984,DM,PirjolSchat}.  For excited baryons, the usual approach consists in organizing states into multiplets  of  the $SU(6) \times O(3)$ group. Even if it has been shown that, for mixed symmetric multiplets, this symmetry is broken at order $\mathcal{O}(N_c^0)$ by spin-orbit interactions, it is a phenomenological fact that these interactions are very small (in the real world with $N_c=3$ they have a magnitude expected for $\mathcal{O}(N_c^{-2})$ effects). Thanks to this observation, the usage of the $SU(6) \times O(3)$ symmetry at leading order is justified. 
 Following this approach,  various works   \cite{DJM95,excitedbaryons,gss,56L2,56L4,MS1,MS2}  have shown  that   the $1/N_c$ expansion is a very useful  tool for  analyzing  the  baryon spectrum. In this work, we assume that the magnitude of spin-orbit interactions is small for highly excited states, \emph{e.g.} for states belonging to $[{\bf 70},5^-]$ and $[{\bf 56},6^+]$ multiplets. Indeed, because of a lack of data, it is not possible to make a detailed study of   these multiplets as it was done in Refs. \cite{excitedbaryons,gss,56L2,56L4,MS1,MS2}  for lower excitations. 

In the $1/N_c$ expansion, the  mass operator for a given $SU(6) \times O(3)$  multiplet is expressed in terms of a series in effective operators \cite{DJM95,excitedbaryons,gss,56L2,56L4,MS1,MS2} ordered in powers of $1/N_c$. 
The coefficients associated with the  operators  are obtained by fitting to the empirical masses.  The various analyses have shown that these coefficients are of natural magnitude or smaller (dynamically suppressed),  lending support to the consistency of the framework. 
To a first approximation,  it turns out that the main features of the spectrum can be 
captured by taking into account a few operators, namely the ${\cal{O}}(N_c)$ spin-flavor singlet operator,
one  ${\cal{O}}(1/N_c)$ hyperfine operator, and the strangeness operator of 
${\cal{O}}(N_c^0 m_s)$. For a few multiplets, the hyperfine $SU(3)$ breaking  $\mathcal{O}(m_s/N_c)$  operator $\hat{S}\cdot \hat{G}_{8}-\frac{1}{2\sqrt{3}}\;\hat{S}^2$  ($\hat{G}_8$ denotes the eighth  component of the axial current, which is one of the $SU(6)$ spin-flavor generators) 
 is  necessary for  achieving  a consistent fit to the empirical masses.    For the finer aspects of the spectrum,  more operators are of course needed.
The coefficients of the  operators considered in this work are ${\cal{O}}(N_c^0)$, and for $SU(3)$ singlet operators the coefficients differ from multiplet to multiplet by amounts ${\cal{O}}(1/N_c)$.
The purpose of this work is to analyze the evolution of  the coefficients as a function of the $O(3)$ quantum number $\ell$.  In particular we focus on the evolution of the  coefficient  associated with the leading spin-flavor singlet operator,   which  determines the Regge trajectories. 

\section{Analysis}

We start by considering the $[{\bf 56},\ell]$ and the  $[{\bf 70},\ell]$ multiplets of  $SU(6) \times O(3)$, which correspond respectively to the symmetric (S) and mixed-symmetric (MS) spin-flavor multiplets
at $N_c=3$.  We entirely disregard possible mixings between these multiplets \cite{GoityMixing}, an approximation that  seems to be consistent phenomenologically as shown by analyses of strong transition amplitudes \cite{strongtrans} as well as electromagnetic transitions \cite{EMtrans}.

For the ground state baryons, which consist of the octet and decuplet in the  $[{\bf 56},0^+]$ multiplet,
the mass formula reads:
\bea
\hat{M}_{\mathrm{GS}} & = & N_c\; c_1 \mathbbm{1}+\frac{1}{N_c} c_{\mathrm{HF}}\left(\hat{S}^2-\frac{3}{4} N_c\right)-c_{\cal{S}}\;\hat{\cal{S}}\nonumber \\
& & +\,\frac{1}{N_c}\, c_{4}\left(\hat{I}^2-\hat{S}^2-\frac{1}{4}\hat{\mathcal{ S}}^2\right),
\eea
where $\hat{S}$, $\hat{I}$ are the baryon spin and isospin operators respectively  and $\hat{\cal{S}}$  is the strangeness operator.  The hyperfine term has been defined   such  that in the limit of a non-relativistic quark picture it corresponds  to the operator $\frac{1}{N_c} \sum_{i\neq j} \vec{s}_i \cdot \vec{s}_j$, {\it i.e.} with the one-body pieces removed. The hyperfine $SU(3)$ breaking operator, mentioned in the introduction,  has been defined in a such way that it  does not  contain terms linear in the strangeness operator   $\hat{\cal{S}}$, and clearly does not contribute to the masses of non-strange ground state baryons. 

For  excited baryons with  $\ell>0$, the hyperfine interaction  of interest can be defined following the   large $N_c$ Hartree picture of the baryon  \cite{witten}:  an excited quark carrying  the orbital angular momentum and a core made out of  the rest $N_c-1$ quarks sitting in the ground state (for $N_c=3$  one can  identify the core with a diquark).  This motivates the choice of hyperfine operator as the one that takes into account the hyperfine interactions between core quarks only.   A second hyperfine operator involves the interaction between core quarks and the excited quark.  In  MS  states  one can separate these two hyperfine interactions explicitly;  it was shown that the latter hyperfine effect is much weaker, and thus  we neglect it here. 
Therefore,    for excited baryons,  except the $[{\bf 56},2^+]$ multiplet, we use the following form for the mass operator:
\bea
\hat{M'}=N_c\,c_1 \mathbbm{1}+\frac{c_{\mathrm{HF}}}{N_c}\! \left(\!{\hat{S}^c\,}^2\!-\!\frac{3}{4} (N_c\!-\!1) \mathbbm{1}\!\right)-c_{\cal{S}}\,\hat{\cal{S}},
\eea
where $\hat{S^c}$ is the spin operator of the core.  Note that the mass formulas generalize beyond the quark model,  as they are entirely given in terms of generators of the spin-flavor group, and thus, only the spin-flavor nature of the states will matter.

For the $[{\bf 56},2^+]$, we add to the mass operator the contribution of the hyperfine $SU(3)$ breaking operator,  which we have modified to be expressed in terms of core operators and to have no term linear in the strangeness of the core:
\bea
\hat{M'} & = & N_c\,c_1 \mathbbm{1}+\frac{c_{\mathrm{HF}}}{N_c}\! \left(\!{\hat{S}^c\,}^2\!-\!\frac{3}{4} (N_c\!-\!1) \mathbbm{1}\!\right)-c_{\cal{S}}\,\hat{\cal{S}} \nonumber \\
& +&  \frac{4\; c_4}{3\, N_c} \, \left( \sqrt{3}\;\hat{S}^c\cdot \hat{G}^{c}_{8} - \frac{1}{2}\;{ \hat{S}^c\,}^2 - \frac{1}{8}\,N_c\; \hat{\mathcal{S}}^c \right).
\eea

For non-strange excited baryons,  the matrix elements of the mass operators in the different cases  are as follows:

\bea
&&\negthinspace\negthinspace M'_{\mathrm{S}}(S)  =  N_c\,  c_1 +\frac{N_c-2}{N_c^2} c_{\mathrm{HF}} \left(S(S+1)-\frac{3}{4} N_c\right),\nonumber\\
&&\negthinspace\negthinspace M'_{\mathrm{MS}}(S\!=\!I)  =  N_c\,  c_1   +\frac{c_{\mathrm{HF}} }{N_c}\nonumber \\ &&~~~~~~~~~~~~~~~~
\times\left(\frac{N_c+2}{N_c}S(S+1)-\frac{3}{4} N_c+\frac{1}{2}\right),\nonumber\\
&&\negthinspace\negthinspace M'_{\mathrm{MS}}(S\!=\!I\!-\!1)  =  N_c \, c_1 \nonumber\\
  &&~~~~~~~~~~~~~~~~~~~~~+\frac{c_{\mathrm{HF}}}{N_c} \left(S(S+2)-\frac{3}{4}(N_c-2)\right),\nonumber\\
&&\negthinspace\negthinspace M'_{\mathrm{MS}}(S\!=\!I\!+\!1) =  N_c \, c_1\nonumber\\&&~~~~~~~~~~~~~~~~~~~~~~~~+\frac{c_{\mathrm{HF}}}{N_c}\! \left(S^2-\frac{3}{4}N_c+\frac{1}{2}\right).
\eea
 For $N_c=3$ the mass formulas become:
\bea
&&\negthinspace\negthinspace N_{\mathrm{GS}}=3\, c_1-\frac{1}{2}\, c_{\mathrm{HF}} ,~ \Delta_{\mathrm{GS}}=3 \, c_1+\frac{1}{2}\,  c_{\mathrm{HF}}, \nonumber\\
&&\negthinspace\negthinspace N_{\mathrm{S}}=3\,  c_1-\frac{1}{6}\,  c_{\mathrm{HF}} ,~ \Delta_{\mathrm{S}}=3\,  c_1+\frac{1}{6}\,  c_{\mathrm{HF}}, \nonumber\\
&&\negthinspace\negthinspace N_{\mathrm{MS}}\left(S=\frac{1}{2}\right)=3 \, c_1-\frac{1}{6}\,  c_{\mathrm{HF}} ,~\\ &&\negthinspace\negthinspace N_{\mathrm{MS}}\left(S=\frac{3}{2}\right)=\Delta_{\mathrm{MS}}\left(S=\frac{1}{2}\right)=3\, c_1+\frac{1}{6}\, c_{\mathrm{HF}}, \nonumber
\eea
where we denote $N\equiv M_N$, etc.   Note that for the MS states we need to specify the total quark spin $S$.   The case of strange baryons where we neglect the $SU(3)$ breaking  hyperfine interaction is obvious, except for the $SU(3)$ singlet $\Lambda$ states in the ${\bf 70}$-plets, where the mass formula becomes:
\beq
\Lambda_{\mathrm{MS}}^{1}=3\, c_1-\frac{1}{2}\, c_{\mathrm{HF}}+ c_{{\cal{S}}}.
\eeq

For the $[{\bf 56},2^+]$,  the matrix elements of the  $SU(3)$ breaking hyperfine operator are lengthy to calculate, and we direct the reader to Refs. \cite{MS2,MS4} for details.

The coefficients $c_1$, $c_{\mathrm{HF}}$, $c_{{\cal{S}}}$ and $c_4$  are determined by fitting to the masses of the corresponding multiplet.  
Tables  1 and 2, for ${\bf 56}$- and  ${\bf 70}$-plets baryons respectively,  display the baryons  listed by the Particle Data Group \cite{PDG06} along with their masses. Some of them ($\ell \leq 4$)  can be identified with a good  level of confidence as belonging  to a definite $SU(6)\times O(3)$ multiplet. For the highest excitations ($\ell=5,6$), the situation is less clear and the identifications proposed are based on Ref. \cite{klempt}.  The Tables also display  the results for the coefficients $c_1$, $c_{\mathrm{HF}}$, $c_{{\cal{S}}}
$ and $c_4$,  and the theoretical masses resulting from the fits. We note here that in the MS states there are  two mixing angles, which  correspond to the mixing of the octet  states with quark spin $S=\frac{1}{2}$ and $\frac{3}{2}$.   In the fit,  these mixings are disregarded because they only originate through  the presence of mass operators we have neglected. We have checked that this approximation does not affect in any  significant way  the conclusions of this work.

In the case of the GS baryons,  as already announced above, the hyperfine $SU(3)$ breaking operator has to be include in the analysis because it  affects  the determination of $c_{\mathrm{HF}}$ through the fit. The result for  $c_{\mathrm{HF}}$ is then consistent with the value  obtained  from the $N$-$\Delta$ mass splitting.   The $\chi^2$ is still large  because of the  $SU(3)$ sub-leading terms that have been disregarded. The inclusion of the higher order  terms  shows the  improvement  expected in the $1/N_c$ expansion \cite{JenkinsLebed}.
  The situation is similar in the $[{\bf 56},2^+]$ multiplet,  where  the hyperfine $SU(3)$ breaking operator has to be included in order to have a consistent fit. One criterion for this consistency is that the values of the coefficients $c_1$, $c_{\mathrm{HF}}$ and $c_{{\cal{S}}}$   are in agreement with the corresponding values obtained in the analysis that includes a  complete basis of operators \cite{56L2}.   
  
  In the  $[{\bf 70}, 1^-]$ multiplet,   the large $\chi^2$ is primarily due to the exclusion of  the spin-orbit operator. That operator   produces  the splitting between the $SU(3)$ singlet $\Lambda$ states, and the failure to describe that splitting gives the main contribution to the   $\chi^2$. This has virtually no effect on the issues we analyze here.  For the {\bf 70}-plets we do not need to include the hyperfine $SU(3)$ breaking term. Note that the available information about the $[{\bf 56}, \ell=4,6]$ and the $[{\bf 70}, \ell=2,3,5]$  states is somewhat limited. In each case,  the information available   is sufficient for determining  the coefficient $c_1$ with enough accuracy for the purpose of  this work,  but the  
  hyperfine and strangeness splittings can be only  roughly determined.  

The main  focus of our study is the  relation across multiplets of the leading order coefficient $c_1$.
Figure 1 shows the plot $(N_c \,c_1)^2$ vs $\ell$.  It displays two distinct   Regge trajectories corresponding to the $[{\bf 56},\ell]$ and the $[{\bf 70},\ell]$ states. In the Hartree picture, the splitting between S and MS trajectories is due to the exchange interaction between the excited quark and the core. Indeed, this exchange interaction turns out to be different for S and MS representations, being order $N_c^0$ in the first case and order $1/N_c$ in the second case. This implies that in large $N_c$ limit there should be two distinct trajectories. The   linear fits to the trajectories  in units of ${\rm GeV}^2$ are as follows\footnote{We considered a fit with a single trajectory, which gives $\chi^2_{\mathrm{dof}}=7.68$, to be  compared to the values 0.57 and 0.06 for the fits to the S and MS trajectories respectively.}:
\bea
\left(3 \;c_1\!\left([{\bf 56}, \ell]\right)\right)^2\!&=&\!(1.179\pm 0.003)+(1.05\pm 0.01)\; \ell,  \nonumber\\
\left(3 \;c_1\!\left([{\bf 70}, \ell]\right)\right)^2\!&=& \!(1.34\pm 0.02)+(1.18\pm 0.02)\; \ell .
\eea
We note that the results for $c_1$ obtained with only non-strange baryons agree, as one would expect, with those obtained including the strange ones.  It is remarkable that the spin-flavor singlet piece of the  squared  masses fit so well on linear Regge trajectories.  The  spread observed   in the Regge trajectories given in terms of the physical masses is, therefore, due to the non-singlet spin-flavor components of the masses, which are dominated by the  hyperfine  components.  For the splitting  between \textbf{56}- and \textbf{70}-plet, the following   linear relation  gives a  fair  approximation:
\bea
&&\negthinspace\negthinspace\left(c_1\! \left([{\bf 56}, \ell]\right)-c_1\! \left([{\bf 70}, \ell]\right)\right)^2 =\nonumber \\  &&~~~~~~~~~~~~~~~~~~~~~~~~~~~~~~ (5.3+4.4\,\ell)\times 10^{-4} \,{\rm GeV^2}.
\eea
This corresponds to a mass splitting    that increases with $\ell$, going from $\sim 70$ MeV at the $\ell=0$ intersect to $\sim 170$ MeV  at  $\ell=6$.
Since hyperfine terms have  this magnitude or larger,  the  differentiation of the  two trajectories can only be clearly seen upon removal of those  terms as we have done here.  One can notice that the identification of the resonance N(2600) as belonging to the  $[{\bf 70},5^-]$ multiplet is well supported by our study. The situation for the N(2700) and $\Delta(2950)$ remains however open.

Note that the quantity with ${\cal{O}}(N_c^0)$ slope    is   $N_c c_1^2$  rather than the one we plotted.
 It is, therefore,  somewhat of a coincidence that at $N_c=3$  the Regge slopes of mesons and of  $N_c^2 c_1^2$ are so similar.  Furthemore, in large $N_c$ limit a  plot linear or  quadratic in $c_1$   would be equivalent, the reason being that the baryon masses are order $N_c$ while the splittings between  multiplets are order $N_c^0$. In the real world,    they differ
slightly, with the quadratic plot giving the best approximation to linear trajectories.

Taking into account the different definition of the hyperfine operator used in this work, which affects the values of $c_1$, we have verified that our results for $c_1$  correspond to those obtained in the analysis Refs. \cite{excitedbaryons,gss,56L2,56L4,MS1,MS2} where complete bases of operators are used. This is a consistency check on the irrelevance of the operators we have neglected for the purpose of our analysis. A similar comment applies to the other coefficients $c_{\mathrm{HF}}$, $c_{\mathcal{S}}$ and $c_4$\footnote{One could make a similar plot to that in Fig. 1 using instead the values of $c_1$ suggested in Refs. \cite{excitedbaryons,gss,56L2,56L4,MS1,MS2}. As presented in Ref. \cite{MS1},  only one Regge trajectory is found in that case. However,  the definition of the bases of operators differs from multiplet to multiplet in Refs. \cite{excitedbaryons,gss,56L2,56L4,MS1,MS2}. This is not the case in present study.}.

It is interesting to notice that the strength of the HF interaction tends to increase with $\ell$.  This  is  shown clearly by  the $[{\bf 70},1^-]$ and   the $[{\bf 56},2^+]$ multiplets,  where the  strength is significantly  larger than for the GS baryons.  Unfortunately,  for baryons with 
 $\ell>2$,  $c_{\mathrm{HF}}$ has large uncertainty and we cannot establish that trend.   According to the $1/N_c$ expansion,  the value of   $c_{\mathrm{HF}}$  differs by  ${\cal{O}}(1/N_c)$ across multiplets,    but in reality  it changes by a factor larger than two in going from the GS  to the $\ell=2$ baryons.  This can be explained by the fact that the hyperfine interaction is more sensitive to the effective size of the core than the other terms in  the  mass formulas.  In particular,  in the quark-diquark picture of the baryon, this sensitivity in the hyperfine  effect   indicates a reduction in the size of the diquark that is significant. 
 The strangeness coefficient $c_{{\mathcal{S}}}$ seems to be bigger for the ground state and the $[{\bf 56},2^+]$ multiplet than for the other cases. We note that the inclusion of the hyperfine $SU(3)$ breaking leads to an enhancement of the fit value of    $c_{{\mathcal{S}}}$.  In the  $[{\bf 70},1^-]$, a more detailed analysis,  including  an additional   $SU(3)$ breaking spin singlet operator \cite{gss},   leads to an enhancement of    $c_{{\cal{S}}}$ as well, bringing it more in line with the values obtained  in the {\bf 56}-plets.
 For other multiplets  the determination of    $c_{{\cal{S}}}$ is rather poor, such as in the  $[{\bf 56},4^+]$ resonance where only one strange baryon is known.  Therefore, it is still possible that  $c_{{\mathcal{S}}}$ has a similar value  across  multiplets, as one would expect.
  Finally, the $c_4$ coefficient, which plays no role in our analysis,  turns out  to have a  large value and  error from the fit to the  $[{\bf 56},2^+]$ multiplet.  A careful consideration of the fit shows that the resonances $\Lambda(1820)$ and $\Sigma(2030)$ play an important role in determining the large value of $c_4$, while the fit gives a poor result for  the mass of $\Lambda(1890)$. The chief difficulty in the $[{\bf 56},2^+]$ multiplet is represented by  the large value of $c_4$,  or equivalently,  the small   masses of  $\Lambda(1820)$ and $\Sigma(2030)$.   It is somewhat puzzling  that these are the only such states in the mass domain, which can  be assigned  to that multiplet.  Although this point is not relevant for this work, it deserves to   be studied more carefully.

\section*{Acknowledgements}

We thank Norberto Scoccola for helpful comments on the manuscript.
This work was supported by DOE (USA) through contract DE-AC05-84ER40150,
by the NSF (USA)   grant  \#~PHY-0300185 (JLG),  by the  I.I.S.N. and the F.N.R.S. (Belgium) (NM).

\newpage

\begin{table*}
\caption{The  coefficients $c_1$, $c_\mathrm{HF}$, $c_\mathcal{S}$, $c_4$ (for the ground state only) and the theoretical masses (MeV) for the {\bf 56}-plets. The experimental masses used for the fit are also presented.}
\begin{center}
 {\scriptsize
\begin{tabular}{l|llcc|ccccc}
\hline \hline
Multiplet \hspace{0.1cm}& Baryon \hspace{0.265cm} &  Name, status \hspace{0.265cm} & Exp. (MeV) &\hspace{0.175cm} Theo (MeV)\hspace{0.175cm} & \hspace{0.24cm}$c_1$ (MeV) & \hspace{0.24cm}$c_{\mathrm{HF}}$ (MeV)  & \hspace{0.24cm} $c_{\mathcal{S}}$ (MeV) &\hspace{0.24cm} $c_4$ (MeV) & \hspace{0.24cm}$\chi^2_{\mathrm{dof}}$ \\
\hline
$[{\bf 56},0^+]$& N$_{1/2}$ & N(939)**** &  $939 \pm 1$ & $939 \pm 2$ & \hspace{0.24cm} $362 \pm 1$ & \hspace{0.24cm} $295 \pm 3$  & \hspace{0.24cm} $208 \pm 3$ & \hspace{0.24cm} $90\pm 5$ & \hspace{0.24cm} 9.1\\
& $\Lambda_{1/2}$ &  $\Lambda(1116)$**** &  $1116 \pm 1$ & $1117\pm 1$ \\
& $^8\Sigma_{1/2}$  &  $\Sigma(1193)$**** & $1192\pm 4$ & $1177 \pm 4$\\
& $^8\Xi_{1/2}$  &  $\Xi(1318)$**** & $1318 \pm 3$ & $1325\pm 4$ \\
& $\Delta_{3/2}$  &  $\Delta(1232)$**** & $1232\pm 1$ & $1233 \pm 2$ \\
& $^{10}\Sigma_{3/2}$ &  $\Sigma(1385)$**** & $1383 \pm 3$ & $1381 \pm 1$ \\
& $^{10}\Xi_{3/2}$  &  $\Xi(1530)$**** & $1532 \pm 1$ & $1529\pm 2$\\
& $\Omega_{3/2}$  &  $\Omega(1672)$**** & $1672 \pm 2$ & $1677 \pm 2$\\
\hline
$[{\bf 56},2^+]$ & N$_{3/2}$  &   N$(1720)$****  &  $1700\pm 50$ & $1682\pm 18$ & \hspace{0.24cm} $603\pm 5$ & \hspace{0.24cm} $767\pm 66$  &\hspace{0.24cm} $233\pm 46$ & \hspace{0.24cm} $416\pm 124$ &\hspace{0.24cm} 1.9\\
& $\Lambda_{3/2}$  &  $\Lambda(1890)$****  &  $1880\pm 30$ & $1822\pm 11$\\
& N$_{5/2}$  &   N$(1680)$**** &  $1683 \pm 8$ & $1682\pm17$\\
& $\Lambda_{5/2}$  &  $\Lambda(1820)$**** & $1820\pm 5$ & $1822\pm11$\\
& $^8\Sigma_{5/2}$  &  $\Sigma(1915)$**** & $1918\pm 18$ & $1915\pm38$\\
& $\Delta_{1/2}$   &   $\Delta(1910)$**** & $1895\pm 25$ & $1938\pm 18$\\
& $\Delta_{3/2}$   &   $\Delta(1920)$*** & $1935\pm 35$ & $1938\pm 18$\\
& $\Delta_{5/2}$   &   $\Delta(1905)$**** & $1895\pm25$ & $1938\pm 18$\\
& $\Delta_{7/2}$   &   $\Delta(1950)$**** &  $1950\pm 10$ & $1938\pm 18$\\
& $^{10}\Sigma_{7/2}$  &  $\Sigma(2030)$**** & $2033\pm 8$ & $2032\pm 18$\\
\hline
$[{\bf 56},4^+]$  & N$_{9/2}$ & N$(2220)$**** & $2245 \pm 65$ & $2245\pm 92$ & \hspace{0.24cm} $770\pm 20$ &\hspace{0.24cm} $398\pm 372$ &\hspace{0.24cm} $110\pm 94$ & &\hspace{0.24cm} 0.13\\
& $\Lambda_{9/2}$ & $\Lambda(2350)$*** & $2355 \pm 15$ & $2355 \pm 21$\\
&  $\Delta_{7/2}$ & $\Delta(2390)$* & $2387\pm 88$ &$2378 \pm 84$\\
& $\Delta_{9/2}$ & $\Delta(2300)$* & $2318\pm 132$ &$2378 \pm 84$\\
& $\Delta_{11/2}$ & $\Delta(2420)$* & $2400\pm 100$ & $2378 \pm 84$\\
\hline
$[{\bf 56},6^+]$  & N$_{13/2}$  & N$(2700)$** & $2806 \pm 207$ & $2806 \pm 207$ &\hspace{0.24cm} $954\pm 40$  &\hspace{0.24cm} $342\pm 720$  \\
&  $\Delta_{15/2}$ & $\Delta(2950)$** & $2920\pm 122$ & $2920\pm 122$\\
\hline
\end{tabular}}
\end{center}
\end{table*}

\begin{table*}
\caption{The  coefficients $c_1$, $c_\mathrm{HF}$, $c_\mathcal{S}$ and the theoretical masses (MeV) for the {\bf 70}-plets. The experimental masses used for the fit are also presented.}
\begin{center}
 {\scriptsize
\begin{tabular}{l|llcc|cccc}
\hline \hline
Multiplet \hspace{0.1cm}& Baryon\hspace{0.5cm} &  Name, status\hspace{0.5cm} & Exp. (MeV)&\hspace{0.25cm} Theo (MeV)\hspace{0.25cm} & \hspace{0.0cm}$c_1$ (MeV) & \hspace{0.0cm}$c_{\mathrm{HF}}$ (MeV) & \hspace{0.25cm}$c_{\mathcal{S}}$ (MeV) & \hspace{0.25cm}$\chi^2_{\mathrm{dof}}$ \\
\hline
$[{\bf 70},1^-]$& N$_{1/2}$ & N$(1535)$**** & $1538\pm 18$ & $1513\pm 14$ &\hspace{0.0cm}$529\pm 5$  & \hspace{0.0cm} $443\pm 19$  & \hspace{0.25cm}$148\pm 13$ & \hspace{0.25cm}61\\
& $^8\Lambda_{1/2}$  &  $\Lambda(1670)$**** & $1670\pm 10$ & $1662\pm 6$\\
& N$_{3/2}$ & N$(1520)$**** & $1523 \pm 8$ & $1513\pm 14$ \\
& $^8\Lambda_{3/2}$ &  $\Lambda(1690)$**** &  $1690 \pm 5$& $1662\pm 6$\\
& $^8\Sigma_{3/2}$  &  $\Sigma(1670)$****  &  $1675 \pm 10$ & $1662\pm 6$\\
& $^8\Xi_{3/2}$   &   $\Xi(1820)$*** & $1823\pm 5$ & $1810\pm15$\\
& N$'_{1/2}$  &  N$(1650)$**** & $1660 \pm 20$ & $1661\pm 17$\\
& $^8\Lambda'_{1/2}$  &  $\Lambda(1800)$*** & $1785 \pm 65$ & $1809\pm 12$ \\
& $^8\Sigma'_{1/2}$  &  $\Sigma(1750)$*** & $1765 \pm 35$ & $1809\pm 12$\\
& N$'_{3/2}$  &   N$(1700)$*** & $1700 \pm 50$ & $1661\pm 17$\\
& N$'_{5/2}$  &   N$(1675)$**** & $1678 \pm 8$ & $1661\pm 17$\\
& $^8\Lambda'_{5/2}$  &  $\Lambda(1830)$**** & $1820\pm 10$& $1809\pm 12$ \\
& $^8\Sigma'_{5/2}$  &  $\Sigma(1775)$**** & $1775\pm 5$& $1809\pm 12$ \\
& $\Delta_{1/2}$   &  $\Delta(1620)$**** &  $1645\pm 30$& $1661\pm 17$\\
& $\Delta_{3/2}$   &  $\Delta(1700)$**** & $1720 \pm 50$& $1661\pm 17$ \\
& $^1\Lambda_{1/2}$  &  $\Lambda(1405)$**** & $1407\pm 4$ & $1514\pm 4$\\
& $^1\Lambda_{3/2}$  &  $\Lambda(1520)$**** & $1520\pm 1$& $1514\pm 4$ \\ 
\hline
$[{\bf 70},2^+]$ & N$'_{1/2}$  &  N$(2100)$* & $1926\pm 26$ & $1987\pm 50$ &  \hspace{0.0cm}$640\pm 16$ & \hspace{0.0cm}$400$ (input) & \hspace{0.25cm}$120\pm 86$ & \hspace{0.25cm}0.03\\
& N$'_{5/2}$  &  N$(2000)$** &  $1981\pm 200$ & $1987\pm 50$\\
& $\Lambda'_{5/2}$  &  $\Lambda(2110)$*** & $2112\pm 40$ & $2108\pm 71$ \\
& N$'_{7/2}$  &  N$(1990)$** & $2016\pm 104$ & $1987\pm 50$\\
& $\Lambda'_{7/2}$  &  $\Lambda(2020)$* & $2094\pm 78$ & $2108\pm 71$\\
& $\Delta_{5/2}$  &  $\Delta(2000)$**  & $1976\pm 237$& $1987\pm 50$  \\
\hline
$[{\bf 70},3^-]$  &  N$_{5/2}$ & N$(2200)$** & $2057\pm 180$ & $2153\pm 67$ & \hspace{0.0cm}$731\pm 17$ & \hspace{0.0cm}$249\pm 315$ & \hspace{0.25cm}$30\pm 159$ &\hspace{0.25cm} 0.15 \\
& N$_{7/2}$  & N$(2190)$**** & $2160\pm 49$ & $2153\pm 67$\\
& N$'_{9/2}$  &  N$(2250)$****  & $2239\pm 76$ & $2236\pm81$ \\
& $\Delta_{7/2}$ & $\Delta(2200)$* & $2232\pm 87$ & $2236\pm 81$\\
& $^1\Lambda_{7/2}$ & $\Lambda(2100)$**** & $2100 \pm 20$ & $2100\pm 28$ \\
\hline
$[{\bf 70},5^-]$  & N$_{11/2}$ & N$(2600)$*** & $2638 \pm 97$ & &\hspace{0.0cm}$900\pm 20$ (Est) \\
\hline
\end{tabular}}
\end{center}
\end{table*}

\begin{figure*}
\begin{center}
 \includegraphics[width=15cm,keepaspectratio]{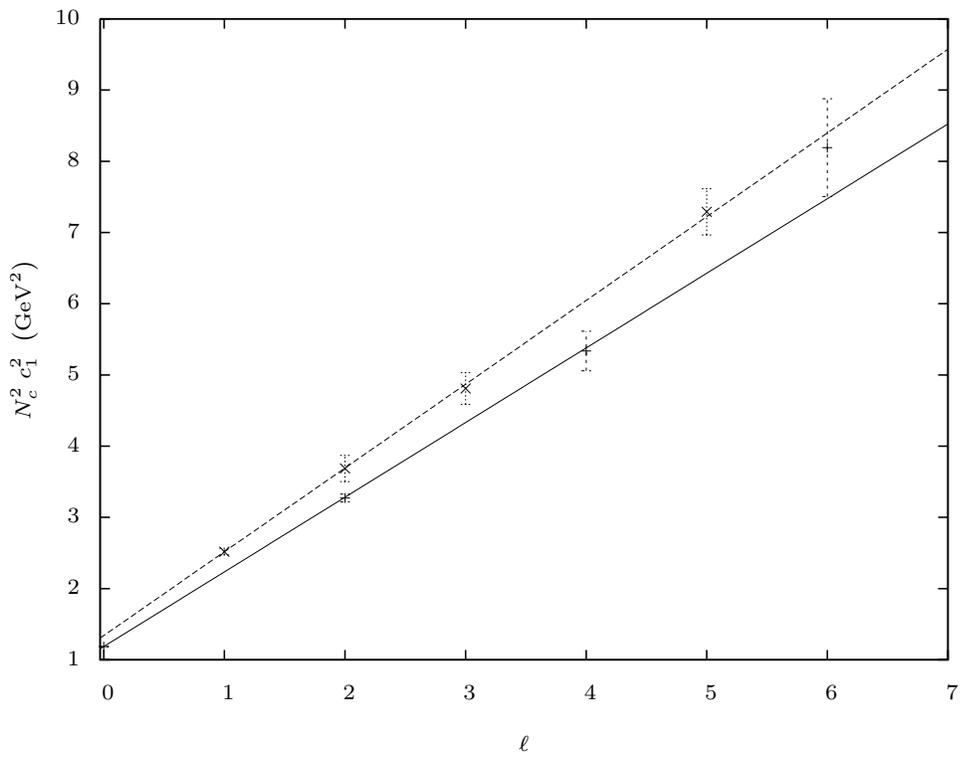}
\end{center}
\caption{Values of the coefficient $(N_c\;c_1)^2$ vs $\ell$ for the {\bf 56}-plets (+) and the {\bf 70}-plets ($\times$). The solid line represents the Regge trajectory for the symmetric multiplets and the dashed line, the Regge trajectory for the mixed symmetric multiplets.}
\end{figure*}

\end{document}